\documentclass[11pt]{article}
\usepackage[a4paper,hmarginratio=1:1,
totalwidth=15.5cm,totalheight=22.9cm]{geometry}
\usepackage{bm,epstopdf,epsfig,amsmath,amssymb,amsfonts,colordvi,latexsym,comment,cancel,
verbatim}
\usepackage{graphicx,graphics}
\usepackage[font=md,captionskip=8pt]{subfig}
\usepackage[usenames,dvipsnames]{color}

\newcommand{\cO}{{\cal O}}

\newcommand{\da}{{\mbox{data}}}

\newcommand{\ra}{\rightarrow}
\newcommand{\be}{\begin{equation}}
\newcommand{\ee}{\end{equation}}
\newcommand{\bea}{\begin{eqnarray}}
\newcommand{\eea}{\end{eqnarray}}

\newcommand{\baa}{\begin{array}}
\newcommand{\eaa}{\end{array}}
\long\def\symbolfootnote[#1]#2{\begingroup
\def\thefootnote{\fnsymbol{footnote}}\footnote[#1]{#2}\endgroup}
\begin{document}

\thispagestyle{empty}
\begin{flushright}
CERN-PH-TH/2012-175\\
\today
\end{flushright}

\vspace{2cm}

\begin{center}
{\Large\bf The  fine-tuning cost of the likelihood in SUSY models.}
\vspace{1cm}

{\bf D. M. Ghilencea$^{\,a,b,\,}$\footnote{E-mail:  dumitru.ghilencea@cern.ch}}
and 
{\bf G. G. Ross$^{a,\,c,\,}$\footnote{E-mail g.ross1@physics.ox.ac.uk}
}

\bigskip

{\small  $^a$ Theory Division, CERN, 1211 Geneva 23, Switzerland.}

{\small $^b$ Theoretical Physics Department, National Institute of Physics}

{\small and Nuclear\, Engineering \, (IFIN-HH)\, Bucharest\, MG-6 077125, Romania.}

{\small $^c$ Rudolf\, Peierls \,Centre \,for Theoretical Physics, University of Oxford,}

{\small 1 Keble Road, Oxford OX1 3NP, United Kingdom.}
\end{center}

\bigskip
\begin{abstract}\noindent
In SUSY models, the  fine tuning of the electroweak (EW) scale  
with respect to their parameters
 $\gamma_i=\{m_0,m_{1/2},\mu_0,A_0,B_0,...\}$ 
and the maximal likelihood $L$ to fit the experimental data  
are usually regarded as two different  problems.
 We show that, if one regards the EW minimum conditions
as constraints that {\it fix} the EW scale, this commonly held view is not 
correct and that  the likelihood 
contains all the information about fine-tuning.
In this case we show that the corrected  likelihood is equal 
to the ratio $L/\Delta$ of the usual likelihood $L$ and
the traditional fine tuning measure $\Delta$ of the EW scale.
A similar result is obtained for  the integrated likelihood over the set 
 $\{\gamma_i\}$, that can   be written as a surface integral 
of the ratio $L/\Delta$, with  the surface in $\gamma_i$ space 
determined by the EW minimum constraints. 
As a result,  a large likelihood 
actually demands a large ratio $L/\Delta$ or equivalently, a  small 
 $\chi^2_{\rm new}=\chi^2_{\rm old}+2\ln\Delta$. 
This shows  the fine-tuning cost to the
likelihood ($\chi^2_{\rm new}$)  of the EW scale stability enforced by SUSY,
that is ignored  in data fits. 
A good $\chi^2_{\rm new}$/d.o.f.$\approx 1$  thus demands SUSY models have 
a fine tuning amount $\Delta\ll \exp({\rm d.o.f.}/2)$, which provides 
a model-independent criterion  for  acceptable fine-tuning. 
If this criterion is not met, one can thus rule out SUSY models without 
a further $\chi^2$/d.o.f.  analysis.
Numerical methods to fit the data  can easily be adapted to  
account for this  effect.
\end{abstract}

\newpage

\section{Fine tuning versus likelihood.}\label{introduction}

There is a commonly held  view  that the fine tuning $\Delta$ 
of the EW scale \cite{Susskind:1978ms}
 wrt UV parameters $\gamma_i=\{m_0,m_{1/2},\mu_0,A_0, B_0,...\}$ 
of SUSY models and the maximal likelihood $L$ to fit the experimental data 
are two separate, unrelated quantities.
The purpose of this letter is to show that  this is not true and that 
there is  actually a mathematical link between the likelihood  and 
the EW scale fine-tuning. 
This is important given  the current LHC SUSY searches and the large
 value of
 $\Delta$ in some models 
\cite{Ghilencea:2012gz,Cassel:2010px},  often seen as a probe against 
SUSY existence \cite{AS}.

We show that, if one regards  the EW minimum conditions as constraints  
that fix the EW scale, the likelihood $L$ or its integrated form \cite{il} over a set 
of parameters $\gamma_i$ include  the effects of the fine-tuning of the EW scale. 
 After eliminating the dependent parameter ($\gamma_\kappa$) fixed  by these constraints, 
we show that the corrected (constrained) likelihood
 is equal  to the ratio of the usual likelihood and the traditional fine 
tuning measure $\Delta$. 
One can also consider  the integrated  likelihood  over a (sub)set 
of $\gamma_i$, and we show that this  can also  be expressed as a surface 
integral of the ratio $L/\Delta$ where $\Delta$ is the 
fine tuning measure (in ``quadrature'')  and the surface in $\gamma_i$ space is 
determined  by the two EW minimum constraints.
Maximizing the  likelihood then requires a large ratio 
$L/\Delta$, usually favouring  minimal fine-tuning  $\Delta$.
Numerical methods to calculate the likelihood in the presence of 
EW minimum constraints can easily be adapted to account for such effects.

We start with some comments about fine tuning, $\Delta$. Common definitions of  $\Delta$ 
are 
\bea\label{tun}
\Delta_{m}=\max\big\vert \Delta_{\gamma_i} \big\vert,
\quad
\Delta_q=\Big(\sum_i \Delta_{\gamma_i}^2\Big)^{1/2},
\quad 
\Delta_{\gamma_i}=\frac{\partial \ln v^2}{\partial \ln \gamma_i^2},\quad
\gamma_i=m_0, m_{1/2}, \mu_0, A_0, B_0,....
\eea
$\Delta_{m}$  was the first measure used  \cite{Ellis:1986yg}.
Its inverse $\Delta_m^{-1}$ was interpreted as a probability of accidental 
cancellations among the  contributions to the EW scale  and its physical implications
for SUSY were previously discussed  in \cite{CS}.
Another measure, $\Delta_q$ also exists, giving for some models
similar numerical results \cite{Ghilencea:2012gz}.
The introduction of these measures  was  based  on physical 
intuition rather than rigorous mathematical grounds. Another drawback is that
$\Delta$ provides a  local measure (in the space 
$\{\gamma_i\}$) of the quantum cancellations that fix the EW scale, while to
compare  models a more  global measure is desirable.
It is often assumed that a solution to the hierarchy problem requires small fine tuning but 
there is no widely accepted upper value for  $\Delta$. 
A related  issue is that one can have  a model 
at a point in the parameter space $\{\gamma_i\}$
 with  very good\footnote{d.o.f.=number of degrees of freedom, equal to
 number of observables fitted minus that
of parameters.}   $\chi^2$/d.o.f. but very large $\Delta$ and another 
point in parameter space with good $\chi^2$/d.o.f. but much smaller $\Delta$. 
In this situation, how do we decide which of these two cases  is better? 
The same question applies when comparing similar  points from different models.

To address this question we note that  to maximize the likelihood 
one adjusts (``tunes'') the  parameters (including $\gamma_i$) to best fit the EW data. 
At the same time, one also must adjust the same parameters
to satisfy the two
EW minimum constraints, one of which actually  {\it fixes} the EW scale vev
(which is central to the definition of fine tuning).
At the technical level there  is no significant distinction
between these constraints and the corresponding adjustment  (``tuning'') of the parameters,
which implies that fine tuning and likelihood must  be related.
 A proper calculation of the  latter 
should then account for fine tuning effects associated with 
EW scale stability under  variations  of $\{\gamma_i\}$, too. 
Here we quantify this effect.

\section{The relation of  constrained likelihood to EW fine tuning.}
\label{th}

Consider the  scalar  potential which in SUSY models has the generic form
\medskip
\begin{eqnarray} 
V&=&  m_1^2\,\,\vert H_1\vert^2
+  m_2^2\,\,\vert H_2\vert^2
- (m_3^2\,\,H_1 \cdot H_2+h.c.)\nonumber\\[3pt]
 &&
 ~+~
(\lambda_1/2) \,\vert H_1\vert^4
+(\lambda_2/2) \,\vert H_2\vert^4
+\lambda_3 \,\vert H_1\vert^2\,\vert H_2\vert^2\,
+\lambda_4\,\vert H_1\cdot H_2 \vert^2\nonumber\\[3pt]
 &&
 ~+~
\Big[\,(\lambda_5/2)\,\,(H_1\cdot  H_2)^2
+\lambda_6\,\,\vert H_1\vert^2\, (H_1 \cdot H_2)+
\lambda_7\,\,\vert H_2 \vert^2\,(H_1 \cdot H_2)+h.c.\Big].
\label{2hdm}
\end{eqnarray}

\medskip\noindent
The couplings $\lambda_j$ and soft masses $m_i$ include
radiative corrections. Introduce the notation
\medskip
\begin{eqnarray}
m^2 &\equiv &
 m_1^2 \, \cos^2 \beta +  m_2^2
 \, \sin^2 \beta - m_3^2 \, \sin 2\beta\nonumber\\[3pt]
\lambda &\equiv &\frac{ \lambda_1^{} }{2} \, \cos^4 \beta 
+ \frac{ \lambda_2^{} }{2} \,  \sin^4 \beta 
+ \frac{ \lambda_{345}^{} }{4} \, \sin^2 2\beta 
+ \sin 2\beta \left( \lambda_6^{} \cos^2 \beta 
+ \lambda_7^{} \sin^2 \beta \right)
\label{ml}
\end{eqnarray}

\medskip\noindent
and  $\lambda_{345}\!=\!\lambda_3\!+\!\lambda_4\!+\!\lambda_5$. 
Using eq.(\ref{ml}), the minimum conditions of $V$ take the compact form
\medskip
\bea
 v^2+\frac{m^2}{\lambda}=0,\qquad\qquad
2 \lambda \frac{\partial m^2}{\partial \beta}-m^2
\frac{\partial\lambda}{\partial\beta}= 0,\qquad
\label{min}
\eea

\medskip\noindent
$v^2\!=\!v_1^2+v_2^2$ is a combination of vev's of
 $h_1^{0}$, $h_2^0$, $\tan\beta\!=\!v_2/v_1$.
The solutions to  constraints (\ref{min})  fix the EW scale  $v$ and $\tan\beta$
for given SUSY UV parameters $\{\gamma_i\}$. It is convenient to define
\medskip
\bea
f_1(\gamma_i; v, \beta, y_t, y_b, \cdots) & \equiv& v-\Big(-\frac{m^2}{\lambda}\Big)^{1/2},
\nonumber\\[8pt]
f_2(\gamma_i; v, \beta, y_t, y_b, \cdots) &\equiv & \tan\beta-\tan\beta_0(\gamma_i,v,y_t,y_b),
\qquad
\gamma_i=\{ m_0, m_{1/2}, \mu_0, A_0, B_0\}.\quad
\label{min2}
\eea

\medskip\noindent
where $\beta_0$ denotes the root of  the second eq in (\ref{min}).
We can therefore use the equations $f_1=f_2=0$ to impose the EW constraints of eqs.(\ref{min})
that relate (fix) the EW scale and $\tan\beta$  in terms of the other  parameters.
The  arguments of $f_{1,2}$ include top, bottom Yukawa couplings, the EW scale $v$ and
 $\tan\beta$ while the dots denote other parameters present at 
one-loop and beyond (gauge couplings, etc), that we ignore without
 loss of generality; $\gamma_i$  shown are for the constrained MSSM, 
but the extension to other SUSY models is trivial. These constraints  
can be assumed to be factorized out of the general likelihood  function 
of a model  $L(\mbox{data}\vert \gamma_i)$, quantifying the likelihood 
  to fit the data with given $\gamma_i$
\medskip
\bea
\label{er}
L\ra L\, \delta\,\big( f_1(\gamma_i; v, \beta, y_t, y_b)\big)\,\,
\delta\,\big(f_2(\gamma_i; v, \beta, y_t, y_b)\big). \,\,\,
\eea

\medskip\noindent
From experiment one also has  accurate measurements such the masses of the
top ($m_t$), bottom ($m_b$)  or  $Z$  boson ($m_Z$). 
To illustrate the point  and to  a good approximation we implement 
 these constraints 
 by Dirac delta functions of  suitable arguments\footnote{One can
implement these via Gaussian distributions, with $\delta_\sigma(x)\ra 1/(\sigma\sqrt{2\pi})
\exp(-x^2/2\sigma^2)$, $\sigma\ra 0$.}, which are again assumed to be
factorized out of the general likelihood function $L$:
\medskip
\bea\label{exp}
L \ra L\, \delta(m_t-m_t^0)\,\delta(m_b-m_b^0) \,\delta(m_Z-m_Z^0),
\eea

\medskip\noindent
where $m_t^0, m_b^0, m_Z^0$ are numerical values from experiment.
The same can be done for other observables, such as the
 well measured $\alpha_{em}$ and $\alpha_3$ 
gauge couplings.

\subsection{The local case.}

When testing a SUSY model with a given set of parameters ($\{\gamma_i\}$),  
one option is to  marginalize  (i.e. integrate) the likelihood $L$ over 
unrelated,   ``nuisance''   parameters that are determined accurately from the data.
An example of such parameters  are the Standard Model 
Yukawa couplings $y_t, y_b,...$ \cite{Cabrera:2008tj}, see 
also \cite{il}\footnote{ 
We integrate over $y_{t,b...}$  instead of the corresponding masses since they
are more fundamental, while masses are derived variables; 
also there is no one-to-one matching of Yukawa to  masses due to $\tan\beta$ dependence.}. 
Another option to eliminate these nuisance parameters ($y_t, y_b,..$) is to construct the 
profile likelihood, in which they are removed by the condition to maximise
$L$ wrt them, for fixed $\{\gamma_i\}$.
In the following we integrate $L$ over  $y_{t,b}$, however,
to ensure our study can be used to  construct 
the profile likelihood, later on 
we  also present the result  without 
integrating over $y_{t,b}$.
Further, we integrate $L$ over 
the  vev $v$ (or well-measured  $m_Z^0$) and $\tan\beta$, 
 which are also fixed by minimization constraints 
(\ref{min}),  (\ref{er}) and (\ref{exp}). One has
 \medskip
\bea
\!\! L(\mbox{data}\vert\gamma_i)\!\! &=&  \!\!\!
N_1\! \int  
\,d (\ln y_t) \, \,d (\ln y_b) \,\, d v\,\,  d(\tan\beta)\,\,
\delta(m_Z-m_Z^0)\,\,\delta(m_t -m_t^0)\,\,\delta(m_b-m_b^0)
\nonumber\\
&\times &\!\!
\delta\,\big(f_1(\gamma_i; v, \beta, y_t, y_b)\big)\,\,
\delta\,\big(f_2(\gamma_i; v, \beta, y_t, y_b)\big)\,
\,\,
L(\mbox{data}\vert \gamma_{i}; v,\beta, y_t,y_b),
\label{opop}
\eea

\medskip\noindent
where $L(\mbox{data}\vert \gamma_i; \beta,v,y_t,y_b)$  is the likelihood to fit the  data
with a particular set of values for   $\gamma_i$, $y_{t,b}$, 
etc, while $L(\mbox{data}\vert\gamma_i)$ is  the (``constrained'') likelihood in the presence of 
the EW constraints and is a function of $\gamma_i$ only; the associated $\chi^2$ is given by 
$\chi^2=-2\ln L$.
In eq.(\ref{opop}) all parameters $\gamma_i, v, \tan\beta, y_t, y_b,\cdots$ 
are independent since the constraints that render them 
dependent variables are enforced by the  delta functions associated to the
theoretical and experimental 
constraints. We integrated over $\ln y_{t,b}$ instead of $y_{t,b}$ for later convenience,
however the conclusion is independent of this.
$N_1$ is  a normalization constant\footnote{
$N_1$   compensates the dimensionful arguments of the three Dirac delta 
functions in the first line of (\ref{opop}).
The integration over $v$ and $\tan\beta$ must be consistent with
the definition of $f_1$ and $f_2$ in (\ref{min2}) in that
$\int dv \delta(f_1(v))=1$ and $\int d(\tan\beta) \delta(f_2(\tan\beta))=1$, so these 
do not generate extra  normalisation factors.
Finally, one could in principle choose to integrate over variables other
than $v,\tan\beta$ (and fixed by the two min conditions), 
but then the functions $f_{1,2}$ should have appropriate
form not to alter their normalisation to unity.}, $N_1=m_b^0\,m_t^0\,m_Z^0$.

To evaluate  $L(\da\vert\gamma_i)$ we use the fact that
 $m_Z={g \, v}/{2}$, $m_t={y_t\, v\,\sin\beta}/{\sqrt 2}$ and
$m_b= {y_t\, v\,\cos\beta}/{\sqrt 2}$
and, after performing the  integrals over $y_t$, $y_b$, $v$, one finds from (\ref{opop})
that
\medskip
\bea\label{twodeltas}
L(\da\vert\gamma_i)&=&
v_0 \int  
d(\tan\beta) \, \,
L\big(\mbox{data}\vert \gamma_i; v_0,\, \beta, \, \tilde y_t(\beta),\, \tilde y_b(\beta)\big),
\nonumber\\[3pt]
&\times &
\delta\big[ f_1\big(\gamma_i;\,v_0,\, \beta,\,\tilde y_t(\beta), \,\tilde y_b(\beta)\big)\big]\,\,
\delta\big[ f_2\big(\gamma_i;\,v_0,\, \beta,\,\tilde y_t(\beta), \,\tilde y_b(\beta)\big)\big]\,\,
\eea

\medskip\noindent
where  $g^2\!\equiv\! g_1^2+g_2^2$ with $g_1$ ($g_2$) 
 the gauge coupling of U(1) (SU(2)) and
\medskip 
\bea\label{stt}
v_0 \equiv  {2 m_Z^0}/{g}=246 \textrm{\,GeV},\qquad 
\tilde y_t(\beta)\equiv \sqrt 2\,m_t^0/(v_0 \sin\beta),\qquad
\tilde y_b(\beta)\equiv \sqrt 2\,m_b^0/(v_0 \cos\beta).
\eea

\medskip\noindent
Note that Yukawa couplings are now functions of $\beta$ only.
Integrating (\ref{twodeltas}) over\footnote{We use
$\delta(g(x))=\delta(x-x_0 ) /\vert g^\prime \big\vert_{x=x_0}$ with 
$g^\prime$ the derivative wrt $x$ evaluated in $x_0$; $x_0$ is the unique root of 
$g(x_0)=0$;  we apply this to a function
 $g(\beta)=f_2(\gamma_i; \beta,v_0, \tilde y_t(\beta), \tilde y_b(\beta)))$ 
for $x\equiv \tan\beta$ with the root $\beta_0=\beta_0(\gamma_i)$.}
$\beta$ gives:
\medskip
\bea
\!
L(\da\vert\gamma_i) 
\!\!\!&=&\!\!\!\!\!
v_0\,\, 
\Big[
L\big(\mbox{data}\vert \gamma_i;\,v_0,\,\beta,\, \tilde y_t(\beta),\, \tilde y_b(\beta)\big)
\,\,\delta\big[ f_1\big(\gamma_i;\, v_0,\,
\beta,\,\tilde y_t(\beta), \,\tilde y_b(\beta)\big)\big]
\Big]_{\beta=\beta_0(\gamma_i)}\,\,
\label{eqeq}
\eea

\medskip\noindent
Here $\beta\!=\!\beta_0(\gamma_i)$  is the unique root of
 $f_2(\gamma_i; v_0, \beta, \tilde y_t(\beta), \tilde y_b(\beta))\!=\!0$, 
via which it becomes a function of $\gamma_i$. It is important to
note that the argument $v$ of $f_1$ was replaced by $v_0$
so the only arguments (variables) of  $f_1$ are $\gamma_i$. 
Further, due to the presence of the delta function, 
one of the $\gamma_i$, hereafter denoted  $\gamma_\kappa$, can be expressed as a function 
of the remaining ones, $\gamma_\kappa\!=\!\gamma_{\kappa}^0(\gamma_j)$, $j\!\not=\!\kappa$, 
where  $f_1$  vanishes if evaluated on 
the set $\{\gamma_j, \gamma_k^0(\gamma_j)\}$, ($j\!\not=\!\kappa$) 
and $L(\mbox{data}\vert\gamma_i)=L(\mbox{data}\vert\gamma_i,i\not\!=\!\kappa)$.
In numerical studies one usually chooses $\gamma_\kappa \!=\! \mu_0$.

The role of the minimum conditions
 in fixing the EW scale is manifest in (\ref{eqeq}) and in a sense one could have
started directly with this eq, but we wanted to show the derivation
of its parametric dependence. 
The effect of $\delta(m_Z-m_Z^0)$ in (\ref{opop}) was to fix the  variable $v$  in (\ref{eqeq}) 
 to the EW scale $v_0=246$ GeV, while the effect of $\delta(m_t-m_t^0)$ 
 and $\delta(m_b-m_b^0)$, upon integration over $y_{t,b}$, was to bring in (\ref{eqeq}) a  
 dependence of $\tilde y_{t,b}$ on $\beta$.

The result of eq.(\ref{eqeq})  has a similar structure 
 in the absence of integrating over Yukawa couplings in (\ref{opop}).
This  is important if one wants to construct  the profile likelihood, 
in which case Yukawa couplings are not integrated out. They 
remain instead independent parameters, on equal footing with $\{\gamma_i\}$,
 and can be adjusted
 to maximise the (unintegrated,  constrained) likelihood, for fixed $\{\gamma_i\}$.
In this case, eq.(\ref{opop}) is modified accordingly and gives
\medskip
\bea
\!\! L(\mbox{data}\vert\gamma_i,y_t, y_b)\!\! &=&  \!\!\!
m_Z^0\! \int  
\, d v\,\,  d(\tan\beta)\,\,
\delta(m_Z-m_Z^0)\,\,
\delta\,\big(f_1(\gamma_i; v, \beta, y_t, y_b)\big)\,\,\nonumber\\[5pt]
&&\,\,\,\times\,\,\,\, \delta\,\big(f_2(\gamma_i; v, \beta, y_t, y_b)\big)\,
\,\, L(\mbox{data}\vert \gamma_{i}; v,\beta, y_t,y_b)
\nonumber\\[5pt]
\!\!\!&=&\!\!
v_0\,\, 
\Big[
L\big(\mbox{data}\vert \gamma_i;\,v_0,\,\beta,\,  y_t,\, y_b\big)
\,\,\delta\big[ f_1\big(\gamma_i;\, v_0,\,
\beta,\, y_t, \, y_b\big)\big]
\Big]_{\beta=\beta_0(\gamma_i; y_t, y_b)},
\label{eqeqp}
\eea

\bigskip\noindent
and the last line above is the counterpart of eq.(\ref{eqeq}).   One can then construct
the profile likelihood from the lhs in (\ref{eqeqp})
by eliminating $y_{t,b}$ via the condition to maximise it
wrt $y_{t,b}$  (for fixed $\{\gamma_i\}$). The profile likelihood is then
$L(\mbox{data}\vert\gamma_i,y_t(\gamma_i), y_b(\gamma_i))$.
The only difference between (\ref{eqeq}), (\ref{eqeqp}) is that the latter
has an extended set of independent parameters to include $y_{t, b}$.
Given this similarity, the results we derive in the following
 are immediately extended to apply to the lhs of (\ref{eqeqp}) 
by a simple replacement $\{\gamma_i\}\ra \{\tilde\gamma_i\}\!\equiv\! \{\gamma_i,y_t, y_b\}$ 
 in the steps below.

Returning to eq.(\ref{eqeq}), we use that $\gamma_\kappa=\gamma_\kappa^0(\gamma_{j\not=\kappa})$ 
is  a  solution to the constraint $f_1=0$,  with independent $\gamma_{j}$ ($j\not=\kappa$)  
{\it fixed} to some numerical values. Then eq.(\ref{eqeq}) can be written as (see footnote 7):
\medskip
\bea\label{op}
L(\da\vert\gamma_i) 
\!\!\!\!&=&\!\!\!\!
 \frac{1}{\Delta_{\gamma_k}}\,\,\delta(\ln\gamma_\kappa-\ln\gamma_\kappa^0(\gamma_j))\,\,
L\big(\mbox{data}\vert \gamma_i;\,v_0,\,\beta,\, 
\tilde y_t(\beta),\, \tilde y_b(\beta)\big)\Big\vert_{\beta=\beta_0(\gamma_i)},\,\,\, j\not=\kappa,
\eea

\medskip\noindent
with\footnote{
With the  arguments of $f_1$ as in 
(\ref{eqeq}), $\tilde v(\gamma_i; \beta)=v_0-f_1=(-m^2/\lambda)^{1/2}$,
 $\partial f_1/\partial\gamma_\kappa
=-\partial \tilde v/\partial\gamma_\kappa$.}: 
\medskip\noindent
\bea\label{sto}
\Delta_{\gamma_\kappa}& \equiv & \bigg\vert
\frac{\partial  \ln \tilde v^2 (\gamma_i;\beta_0(\gamma_i))}{
\partial \ln \gamma_\kappa^2}\bigg\vert_{\gamma_k=\gamma_\kappa^0(\gamma_j);\,j\!\not=\!\kappa};\,\qquad
 \mbox{where}\qquad
\tilde v(\gamma;\,\beta)\equiv \Big(\frac{-m^2}{\lambda}\Big)^{1/2}
\eea

\medskip\noindent
where we used the first eq in (\ref{min2}) and the updated arguments
 of $f_1$ as shown in (\ref{eqeq}).
(For the CMSSM one has $\gamma_j\!=\!\{m_0, m_{1/2}, B_0, A_0\}$, with $\beta$,
$\mu_0$ as output, the latter in the role of $\gamma_{\kappa}$).

Eq.(\ref{op}) is an interesting result that shows that there exists 
a close  relation of the likelihood to the fine tuning $\Delta_{\gamma_{\kappa}}$ 
of the EW scale wrt a parameter ${\gamma_\kappa}$ (c.f. eq.(\ref{tun})).
 The left hand side (lhs) in (\ref{op})  gives the likelihood as a function of the 
{\it independent} variables $\gamma_{j\not=\kappa}$, and is suppressed by the partial 
fine tuning wrt $\gamma_\kappa$ that emerges in denominator on the rhs.
Before discussing this further, note that 
the above effect can also be seen if we formally 
integrate over $\gamma_\kappa$.
 Then
\medskip
\bea\label{rr}
L(\da\vert\gamma_{i\not=\kappa})=\int d(\ln\gamma_\kappa)\,\,L(\da\vert\gamma_i),
\eea
\medskip
\noindent
which gives
%
\bea\label{tt}
L(\da\vert\gamma_{i\not=\kappa})=
\frac{1}{\Delta_{\gamma_\kappa}}\,
L(\da\vert \gamma_i;\,v_0, \beta,\,\tilde y_t(\beta),\tilde y_b(\beta))
\Big\vert_{\beta=\beta_0(\gamma_i); \gamma_\kappa=\gamma_\kappa^0(\gamma_j);\,j\!\not=\!\kappa},
\eea

\medskip
Eq.(\ref{tt})  
shows the relation between the  constrained likelihood (on the lhs)
 and the unconstrained likelihood (on the rhs) without  the EW stability 
constraint. The lhs is what should be maximized when performing data fits  numerically.  
To this purpose one must maximize the ratio of the unconstrained likelihood 
and the fine tuning  $\Delta_{\gamma_\kappa}$ 
wrt  $\gamma_\kappa$ that was eliminated by an EW min condition. 
If  $\Delta_{\gamma_\kappa}\!\gg\! 1$, it reduces considerably the corrected likelihood.

The discussion after eq.(\ref{op}) has assumed that all $\gamma_j$ ($j\not=\kappa$) 
were {\it fixed} and that one can solve  the constraint $f_1=0$ in favour of 
$\gamma_\kappa=\gamma_\kappa^0(\gamma_j)$. One can  avoid these restrictions, 
and obtain the constrained likelihood in a form  manifestly symmetric in all 
$\gamma_i$. Let us denote by $\{\gamma_i^0\}$ (all $i$)  a root of the equation  
that defines the surface of EW minimum $f_1\!=\!0$ where  $f_1$ has the arguments 
displayed in (\ref{eqeq}).  
We denote  $z_i\equiv \ln\gamma_i$, $z_i^0\equiv \ln\gamma_i^0$ and using  a short notation
that only shows  the dependence of $f_1$ on $z_i=\ln\gamma_i$ and after
 a Taylor expansion of $f_1(z_i)$, we have:
$\delta(f_1(z_i))=(1/\vert\nabla f_1\vert)_0\, \delta\big[\vec n.(\vec z-\vec z^0)\big]$,
where $\vec n=(\nabla f_1/\vert\nabla f_1\vert)_0$
 is the normal to this surface,  $\nabla$ is evaluated in basis $\vec z$, 
which  has components $z_1,....,z_n$, 
and the subscript  in $(\nabla f_1)_0$ stands  for evaluation at $z_i^0$ for which $f_1(z_i^0)=0$.
With this, eq.(\ref{eqeq}) can be written 
as 
\medskip
\bea\label{gl2}
L(\da\vert\gamma_i) 
\!\!\!\!&=&\!\!\!\!
 \frac{1}{\Delta_{q}}\,\,\delta\Big(\sum_{j\geq 1} n_j (\ln\gamma_j-\ln\gamma_j^0)\Big)\,\,
L\big(\mbox{data}\vert \gamma_i;\,v_0,\,\beta,\, 
\tilde y_t(\beta),\, \tilde y_b(\beta)\big)\Big\vert_{\beta=\beta_0(\gamma_i)}
\eea

\bigskip\noindent
where $n_j$ are components of $\vec n$ and we used that 
$(1/v_0)\vert \nabla f_1\vert_0=\Delta_q$;  $\Delta_q$ is the fine tuning
in quadrature, defined in (\ref{tun}) with the replacement
 $v\ra \tilde v(\gamma_i,\beta_0(\gamma_i))$ and evaluated at $\gamma_i\!=\!\gamma_i^0$ (all $i$).
 With $n_j$ independent, eq.(\ref{gl2}) is 
satisfied if all $\gamma_i=\gamma_i^0$,  i.e. when $f_1=0$. 
Integrating over  $\ln\gamma_\kappa$ as done in (\ref{rr}), recovers  eq.(\ref{tt}).
Alternatively, one can define a new ``direction'' (variable) $\tilde\gamma$ with
 $\ln\tilde\gamma=\sum_{i\geq 1} n_i\ln\gamma_i$, 
and integrate (\ref{gl2}) over $d(\ln\tilde\gamma)$ instead of $d(\ln\gamma_\kappa)$. 
The result of this integral is
\medskip
\bea\label{ttt}
L({\rm data}\vert \gamma_i^0)=\frac{1}{\Delta_q}\,L({\rm data}\vert \gamma_i;v_0, \beta, 
\tilde y_t(\beta), \tilde y_b(\beta))\Big\vert_{\beta=\beta_0(\gamma_i); \gamma_i=\gamma^0_i}
\eea

\medskip\noindent
This is the counterpart of (\ref{tt}), in a format symmetric over $\gamma_i$
that allows all of them to vary simultaneously on the surface  $f_1=0$.
If all $\gamma_i$ other than $\gamma_\kappa$ are fixed, then (\ref{tt}) is recovered.
To maximize the constrained likelihood, one should actually maximise
the ratio of the unconstrained likelihood and the  fine tuning $\Delta_q$,
evaluated on the surface $f_1=0$.

Eqs.(\ref{tt}), (\ref{ttt}) have an important consequence which is a
change  of the value of $\chi^2$/d.o.f. in
 precision data fits of the models. Let us introduce $\chi^2=-2\ln L$, with similar arguments,
 then from eq.(\ref{tt}),
the corresponding $\chi^2$ values of the
constrained ($\chi^2_{\rm new}$) and unconstrained ($\chi^2_{\rm old}$) likelihoods are related by
\medskip
\bea\label{chisq}
\chi^2_{\rm new}(\gamma_j)=\chi^2_{\rm old}(\gamma_j;\gamma_\kappa^0(\gamma_j))
+2\ln \Delta_{\gamma_\kappa}(\gamma_j),\qquad (j\not=\kappa)
\eea

\medskip\noindent
where we made explicit the arguments of these functions, 
with  $\gamma_\kappa=\gamma_\kappa^0(\gamma_j)$ ($j\not=\kappa$).
Using instead the manifestly symmetric form of eq.(\ref{ttt}), 
the corresponding $\chi^2$ are related by
\bigskip
\bea\label{chisq2}
\chi^2_{\rm new}(\gamma_i)=\Big[\chi^2_{\rm old}(\gamma_i)+2\ln \Delta_q(\gamma_i)\Big]_{f_1=0}
\eea

\medskip\noindent
where the subscript $f_1\!=\!0$ stresses that there is a correlation among the parameters $\{\gamma_i\}$.

The result in  (\ref{chisq2}) and the previous equations
 remain valid if one does not integrate over Yukawa couplings as done in (\ref{opop})
 but uses instead as a starting point  the result of (\ref{eqeqp}).
 All steps after (\ref{eqeqp})
are similar, with the only difference that one must extend the set of arguments $\{\gamma_i\}$ 
of all functions in (\ref{chisq2}) and previous equations, to the set $\{\gamma_i, y_t,y_b\}$. 
In particular $\Delta_q(\gamma_i)$ of eq.(\ref{chisq2}) is replaced by $\Delta_q(\gamma_i, y_t, y_b)$
which  thus includes  the fine tuning  wrt $\gamma_i$ {\it and} $y_t, y_b$ as well, 
all added ``in quadrature'' (i.e. the sum  in the definition of $\Delta_q$ of eq.(\ref{tun}) 
is extended to include $y_t$, $y_b$). This observation is relevant when constructing
the (constrained) profile likelihood function and its associated $\chi^2$. For this,
 Yukawa couplings  $y_{t,b}$   are eliminated from the constrained likelihood  
by the condition of  maximising it wrt each of them, for fixed $\{\gamma_i\}$. 

Eqs.(\ref{chisq}), (\ref{chisq2}) show that $\chi^2_{\rm old}$ 
receives a positive correction  due to  the fine tuning amount of the EW scale.
For a realistic model one must then minimize 
  $\chi^2_{\rm new}$ wrt parameters $\gamma_j$; a good fit requires a
$\chi^2_{\rm new}/{\rm d.o.f.}$ close to one. 
Eq.(\ref{tt}) to (\ref{chisq2}) are the main results
of this paper.

The fine tuning correction can be significant and comparable to $\chi^2_{\rm old}$. For example,
for a modest fine tuning $\Delta=10$, then $2\ln\Delta\sim 4.6$  which is significant,
while  for $\Delta\sim 500$, $2 \ln\Delta\sim 12.5$.
This is the ``hidden'' fine-tuning cost of the likelihood ($\chi^2$) that is ignored 
in current calculations of this quantity in SUSY models, and is associated with 
the EW scale stability that supersymmetry was supposed to enforce in the first place.

Another consequence of the last two equations is that  we can
 infer from them a model-independent value for what is considered acceptable
 fine tuning for any viable model. This is relevant since such
 value was traditionally obtained based  on intuitive rather than 
mathematical grounds.
 With $\chi^2/\mbox{d.o.f.}$ required to be near unity, 
one obtains  the upper bound on the fine tuning, giving
 $\Delta\!~\ll~\!\exp(\mbox{d.o.f.}/2)$. 
This concludes our discussion for  the ``local'' case, 
 without marginalizing over the remaining, independent  parameters.

\subsection{The global case.}

In the following  we  explore a more general case by  computing the 
integrated (i.e. ``global'')  likelihood  over all  $\gamma_i$ parameters.
To compare different SUSY models, that span a similar SUSY space,
the $\gamma_i$-integrated  likelihood can provide useful information \cite{il}. 
This is something very familiar in the Bayesian approach, when computing 
the Bayesian evidence \cite{Cabrera:2008tj}, necessary to compare the relative
probability of different models. The present case of integrated 
likelihood is a special case of the Bayesian case, with the particular choice of log priors.
Integrating over  all  UV  parameters $\gamma_i$, the result for the 
global likelihood to fit the data is
\medskip
\bea
L(\da)& =& 
\int d (\ln\gamma_1)\cdots d(\ln\gamma_n) \,\, L(\da\vert\gamma_i),
\label{margin}
\eea

\medskip\noindent
with $\gamma_i=\{m_0,m_{1/2}, \mu_0,A_0, B_0\}$ for CMSSM.
 For later convenience, we integrate over $\ln\gamma_i$ rather than $\gamma_i$; 
this is one possible choice
of many, and relates to the question whether $\ln\gamma_i$ are more fundamental variables than
$\gamma_i$ or more generally, what the integral measure in $\{\gamma_i\}$ space 
is. 
The rhs of (\ref{margin}) can be converted 
into a surface  integral by using the formula \cite{distro}\footnote{
Eq.(\ref{ly}) can be derived using the discussion around eq.(\ref{gl2}) and the
definition of the element area $dS_{n-1}$.}
\medskip
\bea
\int_{R^n} h(z_1,...,z_n)\, 
\delta(g(z_1,...,z_n))\,d z_1.... d z_n\!=\!\int_{S_{n-1}} d S_{n-1}\, h(z_1,...z_n)\,
\frac{1}{\vert\nabla_{z_i} g\vert}, 
\label{ly}
\eea

\medskip\noindent
The surface $S_{n-1}$ is defined by $g(z_1,...z_n)=0$ and $\nabla$ is the gradient 
in basis $z_i$,
while $dS_{n-1}$ is the element of area on this  surface.
Using (\ref{ly}) with the replacement  $z_i\ra \ln\gamma_i$,
we find from (\ref{eqeq}), (\ref{margin})
\medskip
\bea
L(\da)&=&
\int d (\ln\gamma_1)\cdots d(\ln\gamma_n) \,\, L(\da\vert\gamma_i)
\nonumber\\
&=&
\int_{f_1=0}  d S_{\ln\gamma}\,\,
\frac{1}{\Delta_q(\gamma)}\,\, L\big(\mbox{data}\vert \gamma_i;\,v_0,\,
\beta_0(\gamma_i),\, \tilde y_t(\beta_0(\gamma_i)),\, 
\tilde y_b(\beta_0(\gamma_i))\big)
\label{pofD}
\eea

\medskip\noindent
where
 $d S_{\ln\gamma}$ is the element of area in 
the parameter space $\{\ln\gamma_i\}$.
The last integral is over a surface in $\{\gamma_i\}$ space, given  by 
 EW minimum equation $f_1=0$ (with $f_2=0$ or equivalently
$\beta=\beta_0(\gamma_i)$).
 In the last eq we denoted 
\medskip
\bea
\label{dq}
\Delta_q(\gamma)
&\equiv &
\big\vert \nabla_{\ln\gamma_i}
 \ln \tilde v (\gamma_i; \beta_0(\gamma_i))\big\vert\,\,
=\Big(\sum_{j\geq 1}\,\Delta_{\gamma_j}^2\Big)^{1/2},\quad  
\Delta_{\gamma_j} \equiv  \bigg\vert
\frac{\partial
 \ln \tilde v^2 (\gamma_l;\beta_0(\gamma_l))}{\partial \ln \gamma_j^2}\bigg\vert, 
\eea

\medskip\noindent
with $\gamma_j\equiv m_0, m_{1/2}, \mu_0, A_0, B_0$ and
where   $\nabla_{\ln\gamma_i}$ is the gradient in the coordinate 
space\footnote{Another form of (\ref{pofD}) is found by replacing 
 $dS$, $\nabla$ by their values in $\{\gamma_i\}$  space (instead of $\{\ln\gamma_i\}$)
 and dividing by the  product $\gamma_1....\gamma_n$ under integral (\ref{pofD}).
  $\Delta_{\gamma_j}$  is replaced by  a derivative wrt 
 $\gamma_j$ instead of $\ln\gamma_j$.} $\{\ln\gamma_i\}$
and $\Delta_q(\gamma)$ is the fine tuning in quadrature, $\gamma_i$-dependent.

Eq.(\ref{pofD}) with (\ref{dq})  is a global version 
of  eq.(\ref{ttt}) and shows a similar, interesting result. It presents the
 mathematical origin and the role of the traditional fine tuning measure,
which was not introduced ad-hoc but  turned out to be 
an intrinsic part of the likelihood function
in the presence of the EW minimum constraints.
The result of (\ref{pofD}) is important and useful to compare the
relative probability of two models.

The consequence  of the EW scale stability constraint is that  one should 
maximize in this case 
the integral of the ratio $L/\Delta_q$. Changing the measure of integration 
over $\gamma_i$ also changes the likelihood ``flux''
under the surface integral, but the overall suppression by $\Delta_q$ remains. In 
a Bayesian interpretation, changing the measure under the integral 
corresponds to taking different priors for the variables $\gamma_i$ that were marginalized.
Note however that the emergence of the $1/\Delta_q$ factor under the integral of the
global  $L(\mbox{data})$  (or, more generally, of the global Bayesian evidence)
is entirely an effect of the EW constraints, is {\it independent} of 
the priors and is present  {\it in addition} to the priors factor, 
see  \cite{Ghilencea:2012gz} for further details.
Note that this effect is different from considering a particular
choice for priors proportional to  $1/\Delta_{\gamma_i}$, and also from the so-called 
Jacobian factor, both of which were considered in the past 
in an attempt to account for naturalness \cite{Cabrera:2008tj}. 
For a  Bayesian interpretation of these results and in particular
the integrated likelihood see  \cite{Ghilencea:2012gz,Cabrera:2008tj}.

\bigskip
\subsection{Phenomenological implications.}

The  effects we identified  have phenomenological 
consequences.  Very often, in  SUSY models  a good likelihood
fit to the EW data (i.e. small  $\chi^2$/d.o.f.)  usually prefers  values for the
higgs mass  $m_h$ that have smaller quantum corrections, with $m_h$
near the LEP2 lower bound.
 Further, in models such as the  constrained MSSM (CMSSM), the CMSSM with relaxed,
non-universal higgs soft  masses or the CMSSM with 
non-universal gaugino masses, the fine tuning  grows approximately 
exponentially with the higgs mass,  to large values, of order
 $\Delta\sim 500-1000$  \cite{Ghilencea:2012gz} for  a Higgs mass 
 near the observed value of $\approx 125$ GeV \cite{SMH}.
In such models the constrained
likelihood   $L(\mbox{data}\vert\gamma_i)/\Delta$   is significantly smaller.
This is seen from eqs.(\ref{chisq}), (\ref{chisq2}) showing an increase of $\chi^2$/d.o.f.
by  $2\ln\Delta/\mbox{d.o.f.}$.

For example\footnote{For a recent, detailed discussion
of the $\chi^2$/d.o.f. values in CMSSM and other models  see \cite{HD}. 
In the examples we give we take d.o.f.=9 for the CMSSM, corresponding to 13 observables and 4
parameters.},
in the CMSSM model with $m_h\approx 125$ GeV and taking a
minimal,  optimistic value  $\Delta\approx 100$, then
 $\delta\chi^2/\mbox{d.o.f.} \approx 9.2/9$ which is a very large correction to $\chi^2$,
while taking the value 
$\Delta\approx 500$,
  $\delta\chi^2/\mbox{d.o.f.}\approx 12.5/9$. Even for  models that can have
 $\Delta\approx 10$,  $\delta\chi^2/\mbox{d.o.f.}\approx 4.6/{\rm d.o.f.}$ which
is significant  for  comparable d.o.f.
From this it is clear that the determination of a large fine tuning can rule out models
even before a detailed, traditional $\chi^2$ analysis is undertaken!
In the light of these results, searching for SUSY  models with small 
$\Delta< \cO(10)$ \cite{Cassel:2010px} for the observed value of $m_h$  is well-motivated.

\section{Conclusions}

There exists a commonly held  view that the likelihood 
to fit the data within  a SUSY model and the familiar fine-tuning measure $\Delta$
of the EW scale
 are two distinct problems that should be treated separately.  In this paper we have shown that
this traditional view is not correct and that there is a strong, mathematical link between
these two problems, once the EW minimum conditions are regarded  as constraints 
of the model that fix the EW scale vev. Imposing these constraints, 
the emerging constrained likelihood  $L(\da\vert\gamma_i)$ is proportional 
to the  ratio $L/\Delta$  of the usual, unconstrained likelihood $L$ and the traditional
fine tuning measure $\Delta$.
A similar result applies for  the integrated likelihood over the parameter space.
Maximising the constrained likelihood determines 
how to compare different  points in the parameter space of a model:  
one which has a very good $L$ but 
large $\Delta$,  and another with good $L$ but much smaller $\Delta$.
To conclude, fine tuning is an intrinsic part of the (constrained) likelihood
to fit the EW data, which thus also accounts for the naturalness problem. 

An important consequence  of the above result is 
that the corresponding values of  $\chi^2$  of the constrained
($\chi^2_{\rm new}$) and unconstrained ($\chi^2_{\rm old}$) likelihoods  
are related by the formula {$\chi^2_{\rm new}=\chi^2_{\rm old}+2\ln\Delta$}.
The minimum $\chi^2_{\rm new}$/d.o.f.,  contains a positive correction which has a
negative impact on the overall  data fit of the  model.
This is the ``hidden" fine-tuning cost to the likelihood/$\chi^2$ that is ignored in
current fits of SUSY models, and is associated with the EW scale
stability that low-energy supersymmetry was introduced to enforce.
An acceptable upper bound  of the fine tuning is given by
$\Delta \ll \exp({\rm d.o.f.}/2)$, such that $\chi^2_{\rm new}$/d.o.f. is 
not significantly worse. Generically this requires  $\Delta<\cO(10)$.
If this bound is not respected, this analysis shows that one can  rule out a
 model  without a detailed  $\chi^2$ analysis.

In conclusion, we have shown that the EW scale  fine tuning does play a role in establishing
if a model is realistic or  not in a probabilistic sense.
Current  numerical methods to fit the data can easily be adapted to   
account  for this effect.

\section{Acknowledgements}

D.M.G. thanks  S. Kraml (CNRS) and B.~Allanach (Cambridge) 
for many useful discussions.

\end{document}